\documentclass[a4paper,11pt]{article}
\pdfoutput=1
\usepackage[utf8]{inputenc}
\usepackage[T1]{fontenc}
\usepackage{jheppub}
\usepackage{natbib}

\usepackage[usenames,dvipsnames,svgnames,table,x11names]{xcolor}
\usepackage{lmodern,amsmath,multirow,cancel}
\usepackage{booktabs}
\usepackage{xstring}
\usepackage{ifthen}

\usepackage{ascmac,braket,bm,mathrsfs,amsthm,amsfonts}
\usepackage{hyperref}
\usepackage{graphicx}
\usepackage{subcaption} % Allow for subfigures
\usepackage{comment}
\usepackage[normalem]{ulem}

\usepackage{titlesec}
\newcommand*{\justifyheading}{\raggedright}
\titleformat{\chapter}[display]
  {\normalfont\huge\bfseries\justifyheading}{\chaptertitlename\ \thechapter}
  {20pt}{\Huge}
\titleformat{\section}
  {\normalfont\Large\bfseries\justifyheading}{\thesection}{1em}{}
\titleformat{\subsection}
  {\normalfont\large\bfseries\justifyheading}{\thesubsection}{1em}{}
\titleformat{\subsubsection}
  {\normalfont\bfseries\justifyheading}{\thesubsubsection}{1em}{}
\usepackage[english]{babel}
\usepackage[dvipsnames]{xcolor}
\numberwithin{equation}{section}

\renewcommand{\star}{\ast}

\usepackage{slashed}
\newcommand{\Slash}[1]{{\ooalign{\hfil/\hfil\crcr$#1$}}}

\newcommand{\eg}{{\em e.g.}}

\newcommand{\gsim}{ \mathop{}_{\textstyle \sim}^{\textstyle >} }
\newcommand{\lsim}{ \mathop{}_{\textstyle \sim}^{\textstyle <} }
\usepackage{adjustbox}
\usepackage{tabularx}
\newcolumntype{Y}{>{\centering\arraybackslash}X} %for tabularx
\usepackage{booktabs} %for \toprule
\usepackage{colortbl} %for \rowcolors
\def\beq#1\eeq{\begin{align}#1\end{align}}
\RequirePackage{xspace}
\def\Bbar    {\kern 0.18em\overline{\kern -0.18em B}{}\xspace}

\preprint{IPMU23-0042}

\title{Closer look at the matching condition for
radiative QCD \boldmath{$\theta$} parameter}

\author[a]{Tatsuya Banno,}
\author[a,b,c]{Junji Hisano,}
\author[b,d]{Teppei Kitahara,}
\author[a]{and Naohiro Osamura}

\affiliation[a]{
  Department of Physics, Nagoya University, Furo-cho Chikusa-ku, Nagoya 464-8602 Japan
}
\affiliation[b]{
  Kobayashi-Maskawa Institute for the Origin of Particles and the
  Universe, Nagoya University,
  Furo-cho Chikusa-ku, Nagoya 464-8602 Japan
}
\affiliation[c]{
  Kavli IPMU (WPI), UTIAS, The University of Tokyo, Kashiwa 277-8584, Japan
}

\affiliation[d]{
  CAS Key Laboratory of Theoretical Physics, Institute of Theoretical Physics, Chinese Academy of Sciences, Beijing 100190, China}

\emailAdd{banno.tatsuya.p8@s.mail.nagoya-u.ac.jp}
\emailAdd{hisano@eken.phys.nagoya-u.ac.jp}
\emailAdd{teppeik@itp.ac.cn}
\emailAdd{osamura.naohiro.j2@s.mail.nagoya-u.ac.jp}

\abstract{In this paper, we scrutinize a radiatively generated QCD $\theta$ parameter at the two-loop level 
based on both full analytical loop functions with the Fock-Schwinger gauge method and the effective field theory approach, using simplified models.
We observe that the radiatively generated $\theta$ parameters at the low energy scale precisely match  
between them. 
It provides validity 
to perturbative loop calculations of the QCD $\theta$ parameter with the Fock-Schwinger gauge method.
Furthermore, it is also shown that the
ordinary Fujikawa method for the radiative $\theta$ parameter by using  $\bar\theta = - \text{arg}\,\text{det}\,\mathcal{M}_q^{\rm loop}$ 
does not cover all contributions in the simplified models. 
But, we also find that when there is a scale hierarchy in $CP$-violating sector, evaluation of the Fujikawa method is numerically sufficient.
As an application,
we calculate the radiative $\theta$ parameter at the two-loop level
in a slightly extended Nelson-Barr model,  where 
the spontaneous $CP$ violation occurs to solve the strong $CP$ problem. 
It is found a part of the radiative $\theta$ parameters cannot be described by the Fujikawa method.}

\keywords{$CP$ violation, Electric Dipole Moments}

\begin{document}
\sloppy %https://tex.stackexchange.com/questions/9107/how-can-i-make-my-text-never-go-over-the-right-margin-by-always-hyphenating-or-b

% get rid of JHEP header
\makeatletter\renewcommand{\@fpheader}{\ }\makeatother

\maketitle

\renewcommand{\thefootnote}{\#\arabic{footnote}}
\setcounter{footnote}{0}

%=======================================================
%        INTRODUCTION
%=======================================================
\section{Introduction}

In the renormalizable QCD Lagrangian, the following terms are $P$-odd and $T$-odd, and then $CP$-odd terms under the $CPT$ invariance,
\begin{align}
    \mathcal{L}_{\Slash{P},\,\Slash{T}}
    = - \sum_{q=\text{all}} \text{Im}(m_q) \bar{q} i \gamma_5 q + \theta_G \frac{\alpha_s}{8\pi } G_{\mu\nu}^a\tilde{G}^{a\mu\nu}\,.
    \label{eq:thetaG}
\end{align}
Here, $m_q$ stands for the complex quark masses with $m_q \equiv |m_q|\exp({i\theta_q})$,  $G_{\mu\nu}^a$ is the gluon field-strength tensor, 
$\tilde{G}^{a\mu\nu} \equiv \frac{1}{2} \epsilon^{\mu \nu \rho \sigma} G^a_{\rho \sigma}$ with $\epsilon^{0123} = +1$, 
and $\alpha_s =g_s^2/(4\pi)$ is the $SU(3)_C$ coupling constant. It is known that one can turn off the first terms with the axial rotation of quarks in order to get the physical QCD $\theta$ parameter,  $\bar{\theta}\,(\equiv \theta_G-\sum_q \theta_q)$, defined as 
\begin{align}
    \mathcal{L}_{\Slash{P},\,\Slash{T}}
    =  \bar{\theta} \, \frac{\alpha_s}{8\pi } G_{\mu\nu}^a\tilde{G}^{a\mu\nu}\,.
    \label{eq:theta}
\end{align}
This QCD $\theta$ parameter would be expected to be $O(1)$ naturally while it is constrained by the neutron electric dipole moments (EDM) measurements as $|\bar{\theta}|\lesssim 10^{-10}$ \cite{Abel:2020pzs,Liang:2023jfj}.\footnote{%
Recently, EDMs of paramagnetic molecules such as ThO\cite{ACME:2018yjb} or HfF$^+$\cite{Roussy:2022cmp} are significantly improved, 
and the constraint on the QCD $\theta$ parameter from $CP$-odd semileptonic interactions
by these experiments reaches $|\bar{\theta}| < 10^{-8}$~\cite{Flambaum:2019ejc} if the source of $CP$-violation is dominated by the hadronic sector.}
This serious naturalness problem is called the strong $CP$ problem. 

Some models have been proposed to solve the strong $CP$ problem. One of them is the Peccei-Quinn (PQ) mechanism \cite{Peccei:1977hh,Weinberg:1977ma,Wilczek:1977pj}. The mechanism works well to solve the problem since the strong dynamics tunes the QCD $\theta$ parameter to be zero with the axion field, 
while an introduction of the PQ global symmetry suffers from the quality problem of the global symmetry in quantum gravity \cite{Kamionkowski:1992mf,Holman:1992us,Barr:1992qq}.
The second option is to forbid the QCD $\theta$ term with the $CP$ \cite{Nelson:1983zb,Barr:1984qx,Barr:1984fh} or $P$ symmetry \cite{Beg:1978mt,Mohapatra:1978fy,Babu:1989rb, Barr:1991qx}.
Those discrete symmetries are broken in the Standard Model (SM),
so that the extensions of the SM with those symmetries are mandatory.
In the left-right symmetric extension of the SM, the generalized $P$ symmetry forbids the QCD $\theta$ term (and also the imaginary parts of quark masses) 
at the tree level. 
After the spontaneous breaking of the $P$ symmetry to match the SM, the QCD $\theta$ parameter is radiatively generated.
Also, it is known that there is no quality problem in the mechanism \cite{Berezhiani:1992pq,Craig:2020bnv}.
On the other hand, 
in the Nelson-Barr models,
the $CP$ symmetry is imposed so that the QCD $\theta$ term is forbidden at the tree level. Then,
the $CP$ symmetry is spontaneously broken to generate the $CP$ phase in the CKM matrix, and the model is tuned to keep the QCD $\theta$ parameter zero at the tree level, but it is generated radiatively.

In those models of the second option to solve the strong $CP$ problem, the QCD $\theta$ parameter is radiatively generated and a finite neutron EDM is predicted even if the energy scale of the symmetry breaking is much higher than the weak scale \cite{Hall:2018let}.
Thus, the precise evaluation of the radiative $\theta$ parameter 
is important to judge whether the models are viable under the current constraint from the neutron EDM measurements and also whether they are testable in future EDM experiments for the neutron and other developed systems.
In past literature, \eg, Refs.~\cite{Babu:1989rb,Hall:2018let,Craig:2020bnv,deVries:2021pzl},
the radiative corrections to the QCD $\theta$ parameter have been evaluated by using the radiative corrections to the imaginary parts of the colored particle masses \cite{Ellis:1978hq},
\beq
\bar{\theta} = - \textrm{arg}\,\textrm{det}\,\mathcal{M}_q^{\rm loop}\,,
\label{eq:argdetm}
\eeq
where $\mathcal{M}_q^{\rm loop}$ is the renormalized (up- and down-type) colored fermion mass matrix.
This relation is based on
the Fujikawa method \cite{Fujikawa:1979ay} 
or the Adler-Bell-Jackiw anomaly with Adler-Bardeen theorem \cite{Adler:1969gk,Bell:1969ts,Adler:1969er,Fujikawa:1979ay,tHooft:1972tcz}.

However, it has been shown in Refs.~\cite{Khriplovich:1985jr,Pospelov:1994uf}, in the context of the SM,  
that the radiative correction to the QCD $\theta$ parameter can be evaluated directly and perturbatively,
not via the imaginary parts of quark masses \cite{Ellis:1978hq},  using an external field technique, while details of the technique are not written.
In Ref.~\cite{Hisano:2023izx}, it was shown that  such a radiative $ \theta$ parameter can be evaluated by the Feynman diagrammatic calculations
 using the Fock-Schwinger gauge method \cite{Fock:1937dy,Schwinger:1951nm,key86364857,Cronstrom:1980hj,Shifman:1980ui,Dubovikov:1981bf,Novikov:1983gd}.
The QCD $\theta$ term is equivalent to a total derivative term, 
and hence the radiative correction to the QCD $\theta$ parameter is not perturbatively generated if the translational invariance is maintained.
The operator-Schwinger method \cite{Novikov:1983gd} using the log-determinant of the Dirac operators 
works for the evaluation of the radiative corrections to the action with keeping the translational symmetry.
However, the operator-Schwinger method makes it difficult to go beyond the one-loop corrections to the QCD $\theta$ parameter. 
On the other hand, the Fock-Schwinger gauge method is applicable to the higher-loop corrections \cite{Abe:2017sam,Hisano:2023izx}. 
In fact, the Fock-Schwinger gauge violates the translational invariance, while it is shown that the invariance is recovered in the physical overvalues \cite{Nikolaev:1981ff,Nikolaev:1982rq}.
Some of the authors applied the Fock-Schwinger gauge method for evaluation of the QCD $\theta$ parameter in the minimal left-right model \cite{Hisano:2023izx}. 
It is found that the two-loop contribution to the QCD $\theta$ parameter vanishes in the model. 
The order of magnitude of the three-loop contribution from the diagrams is evaluated 
and it is found that it can be as large as the current experimental bound of the neutron EDM measurement. 

In this paper, we derive the structure of the radiative correction to the QCD $\theta$ parameter at two-loop level. 
We compare the full two-loop corrections with those evaluated by the Fujikawa method taking into account only the radiative correction to the colored fermion masses at one-loop level in Eq.~\eqref{eq:argdetm}. 
In Sec.~\ref{sec:1loop}, we review several calculations related to the QCD $\theta$ parameter at one-loop level by using $CP$-violating low-energy effective theory.
In Sec.~\ref{sec:2loop}, we study minimal models at two-loop level,
in which colored fermion(s) and a real scalar have $CP$-violating Yukawa coupling(s) inducing the radiative QCD $\theta$ parameter.
We find that if the colored fermions and the real scalar have masses comparable to each others, the Fujikawa method in Eq.~\eqref{eq:argdetm} does not cover full contribution and we need to calculate the two-loop contribution to the QCD $\theta$ parameter directly.

In Sec.~\ref{sec:NB}, we discuss the two-loop corrections to the QCD $\theta$ parameter in an extension of the Nelson-Barr model, as a concrete example of UV model. 
In the model, we introduce the four-point couplings of the SM Higgs ($H$) and the spontaneous $CP$-violating scalar ($\Sigma_a$, $a=1,\cdots, N_\Sigma$), 
and a real scalar ($S$)  fields,  responsible to the real vector-like fermion mass. Both interactions of $|H|^2\Sigma_a\Sigma_b^\star$ and $S^2 \Sigma_a\Sigma_b^\star$, induce the QCD $\theta$ parameter at two-loop level. 
The former is determined by the radiative correction to the SM quark masses. On the other hand, the latter depends on the heavy particle mass spectrum, and it may not be determined by the radiative correction to the fermion mass.
Section \ref{sec:conclusion} 
is devoted to conclusions.
Details of the two-loop calculations
are given in the Appendix.

\section{One-loop calculations for matching condition of QCD \texorpdfstring{\boldmath{$\theta$}}{theta}  parameter}
\label{sec:1loop}

In this section, 
we evaluate the QCD $\theta$ parameter by integrating out a colored fermion $q$, 
whose mass and interactions are $CP$ violating, 
in low-energy effective field theories at one-loop level. 
The general effective Lagrangian of the colored fermion is given in QCD up to dimension-five operators as 
\beq
{\cal L}_{\rm eff} &=
\bar{q}\left[i \Slash{D}- \left(m_q^\star P_L +
m_q P_R\right)\right]q
-\frac{1}{2} g_s {\mu}_q\bar{q}
\left(\sigma \cdot G \right)  q 
-\frac{i}{2} g_s {d}_q\bar{q} \left(\sigma \cdot G \right) \gamma_5 q\,,
\label{EFTF}
\eeq
where 
$D_\mu=\partial_\mu+ig_s T^a G_\mu^a$, 
$T_a = \frac{1}{2}\lambda_a$ ($\lambda_a$ are the Gell-Mann matrices),
$(\sigma \cdot G) = \sigma^{\mu\nu}T^a G_{\mu\nu}^a$, 
$\sigma_{\mu \nu} =
\frac{i}{2} [\gamma_\mu, \gamma_\nu]$,
and
$P_{L/R} = (1 \mp \gamma_5)/2$.
If the $CP$-violating phase of the colored fermion mass, $m_q\equiv |m_q|\exp(i\theta_q)$, is removed by the axial rotation (chiral rotation) $q \to \exp(- \frac{i}{2} \theta_q \gamma_5)q$, one obtains \cite{Adler:1969gk,Bell:1969ts,Adler:1969er,Fujikawa:1979ay,tHooft:1972tcz}
\beq
\mathcal{L}_{\rm eff} = \bar{q}\left(i \Slash{D}- |m_q|\right)q
-\frac{1}{2} g_s \tilde{\mu}_q\bar{q}
\left(\sigma \cdot G \right)  q 
-\frac{i}{2} g_s \tilde{d}_q\bar{q} \left(\sigma \cdot G \right) \gamma_5 q
-\theta_q \frac{\alpha_s}{8 \pi} G^a_{\mu\nu}\tilde{G}^{a\mu\nu}\,,
\label{EFTF2}
\eeq
with
\beq
\tilde{\mu}_q = \cos (\theta_q) \mu_q + \sin (\theta_q) d_q\,, \quad 
\tilde{d}_q = - \sin (\theta_q) \mu_q + \cos (\theta_q) d_q\,,
\label{eq:physicaldq}
\eeq
where $\tilde{\mu}_q$ and $\tilde{d}_q$ correspond to the (anomalous) chromo-magnetic dipole moment (CMDM) and chromo-electric dipole moment (CEDM).\footnote{%
Note that a nonrelativistic term of the CMDM is included in the Dirac operator.}
Here, we ignore the dimension-six four-Fermi operators.\footnote{%
In this paper, we also discarded the QED interactions.} 
They are mixed with the dipole moment operators at one-loop level renormalization-group evolution
if they are present \cite{Hisano:2012cc}. 
The renormalization-group evolution effect will be taken into account in the next section.

We integrate out the colored fermion with the operator Schwinger method to evaluate the QCD $\theta$ parameter at one-loop level.
The method is transparent compared with the Fock-Schwinger gauge method when considering the one-loop corrections \cite{Hisano:2023izx,Novikov:1983gd}.
They must be coincident with each other when we calculate the gauge invariant quantities. 

First, we consider the contribution of the colored fermion mass phase to the QCD $\theta$ parameter. 
It is shown in Ref.~\cite{Hisano:2023izx} that 
in the operator-Schwinger method, the one-loop correction to the effective action $\Delta S$ after integrating out the colored fermion field is 
given as 
\begin{eqnarray}
\Delta S &=&-i \operatorname{Tr} \log\left|\left| \Slash{P}-(m_q^\star P_L +
m_q P_R)\right|\right|\,,
\end{eqnarray}
where $P_\mu\equiv i D_\mu$.
The derivative of $\Delta S$ over $m_q$ for $P_R$ is 
\cite{Hisano:2023izx,Novikov:1983gd}
\beq
\frac{d}{d m_q}\Delta S
&=
i\operatorname{Tr}\left[
\frac{1}{P^2-|m_q|^2-\frac12 g_s(\sigma \cdot G)} m_q^\star P_R
\right]\,,
\label{dsdm}
\eeq
where
 $
\left[P_\mu, P_\nu\right]= - i g_s G_{\mu \nu}
 $
is used. 
Since the $G \tilde{G}$ term includes the Levi-Civita tensor,
the QCD $\theta$ term comes from the second order of the perturbation of $\frac{1}{2}g_s(\sigma \cdot G)$ in the denominator in the right-handed side of Eq.~\eqref{dsdm},
\beq
\frac{d}{d m_q}\Delta S 
&\supset
i\operatorname{Tr}\left[
\frac1{P^2-|m_q|^2}\left(\frac12 g_s (\sigma \cdot G)\right)\frac1{P^2-|m_q|^2}\left(\frac12 g_s (\sigma \cdot G)\right)
m_q^\star P_R\right]\nonumber\\
&=
\int d^4 x\int \frac{d^4p}{(2\pi)^4} 
\frac{g_s^2}2\frac{-m_q^\star}{(p^2-|m_q|^2)^3}G^a_{\mu\nu} \tilde{G}^{a\mu\nu}+\cdots\nonumber\\
&= i\int d^4x \frac{g_s^2}{64\pi^2}\frac1{m_q}
G^a_{\mu\nu} \tilde{G}^{a\mu\nu}+\cdots\,,
\eeq
where ${\rm Tr}(T^aT^b)=(1/2)\delta^{ab}$ is used. 
Here, $P^2$ is replaced with $(i\partial)^2$, and it is integrated in the momentum space. 
Similarly, $d\Delta S/dm_q^\star$ leads to the $G \tilde{G}$ term. By integrating $d\Delta S/dm_q$ with $m_q$ and $d\Delta S/dm_q^\star$ with $m_q^\star$, we get 
\beq
\Delta {\cal L} &=-i\frac{\alpha_s}{16\pi}\log\frac{m_q^\star}{m_q} G^a_{\mu\nu} \tilde{G}^{a\mu\nu}
\nonumber\\
&=
-\theta_q\frac{\alpha_s}{8\pi} G^a_{\mu\nu} \tilde{G}^{a\mu\nu}\,.
\eeq
As expected, the integration of the colored fermion gives the contribution to QCD $\theta$ term, consistent with the Fujikawa method in Eq.~\eqref{EFTF2}. 

Next, we show the contribution of the dipole moments, $\tilde{\mu}_q$ and $\tilde{d}_q$, in Eq.~{\eqref{EFTF2}. The CEDM is $T$-odd and $P$-odd, and it also contributes to the QCD $\theta$ parameter as shown below. 
When the phase of the colored fermion mass is removed by the chiral rotation,
the physical CEDM $\tilde{d}_q$ is given by
$\tilde{d}_q=\cos(\theta_q) d_q - \sin(\theta_q) \mu_q$ in Eq.~\eqref{eq:physicaldq}. The evaluation of the physical CEDM contribution to the QCD $\theta$ parameter
is more straightforward in the operator-Schwinger method. The one-loop correction to the effective action after integrating out the colored fermion field is 
\beq
\Delta S&=
-i \operatorname{Tr} \log\left|\left| \Slash{P}-|m_q|
-\frac{i}2 g_s \tilde{d}_q (\sigma \cdot G)\gamma_5 
\right|\right| \,\nonumber\\
&\supset
-i \operatorname{Tr}
\left[
\frac1{P^2-|m_q|^2-\frac12 g_s(\sigma\cdot G)}
|m_q|
\left(-\frac{i}2 g_s \tilde{d}_q (\sigma \cdot G)\gamma_5\right)
\right]\,\nonumber\\
&=
-
|m_q|\tilde{d}_q  \operatorname{Tr}\left[
\frac1{P^2-|m_q|^2}\left(\frac12 g_s(\sigma\cdot G)\right)\frac1{P^2-|m_q|^2}
\left(\frac12 g_s(\sigma \cdot G)\right)\gamma_5
\right]\cdots \,\nonumber\\
&=
2 |m_q|\tilde{d}_q   \times 
(16\pi^2\mu^{2\epsilon})\int \frac{d^dp}{(2\pi)^d}\frac1{(p^2-|m_q|^2)^2}\times 
\int d^4x \frac{\alpha_s}{8\pi} G^a_{\mu\nu} \tilde{G}^{a\mu\nu} \nonumber\\
&=
2 |m_q|\tilde{d}_q
\times 
\left(\frac{1}{\bar{\epsilon}}-\log\frac{|m_q|^2}{\mu^2}\right)
\times 
\int d^4x \frac{\alpha_s}{8\pi} G^a_{\mu\nu} \tilde{G}^{a\mu\nu}\,. 
\eeq
In the second line, we keep the leading term of the CEDM in the expansion. In the fourth line, $P^2$ is replaced by $(i\partial)^2$, and it is integrated out in the momentum space again.
Since the integral includes UV divergence, we introduce the dimensional regularization $d=4-2\epsilon$ with the renormalization scale $\mu$.
If the $\overline{\text{MS}}$ scheme is adopted,  
$1/\bar{\epsilon}\,(\equiv 1/\epsilon -\gamma_E+\log 4\pi$ and $\gamma_E$ is the Euler's constant)
is subtracted by the counter term. It is found that the QCD $\theta$ parameter depends on the renormalization scale as \cite{Jenkins:2017dyc}
\beq
\mu \frac{d \theta}{d\mu} &=
4 |m_q|\tilde{d}_q\,,
\label{eq:RGE}
\eeq
and the threshold correction at the colored fermion mass ($\mu=|m_q|$) due to the CEDM vanishes. 

It is known that the CEDMs of the colored fermions generate the $T$-odd and $P$-odd higher-dimensional gluon operators, such as 
the Weinberg operator with dimension six, $G^2\tilde{G}$, in addition to the QCD $\theta$ parameter when the colored fermions are integrated out \cite{Boyd:1990bx,Braaten:1990gq,Chang:1990jv,Dine:1990pf,Abe:2017sam}.
On the other hand, the colored fermion mass phases induce only the QCD $\theta$ parameter, but not the higher-dimensional gluon operators. 
The axial rotations of the colored fermions turn off the colored fermion mass phases and simultaneously generate the QCD $\theta$ term in Eq.~\eqref{EFTF2}. This can be derived with the Fujikawa method \cite{Fujikawa:1979ay} or the explicit calculation of the above one-loop colored fermion diagram \cite{Hisano:2023izx}. Now, we consider whether $T$-odd and $P$-odd higher-dimensional gluon operators, such as the Weinberg operator, are generated or not, when the quark mass phases are non vanishing.

In the Fujikawa method, the higher-dimensional gluon operators are not generated from the Jacobian of the chiral transformation \cite{Fujikawa:1979ay}. 
However, an UV cut-off parameter has to be introduced in order to define the anomalous Jacobian, which includes only the QCD $\theta$ term in the limit of the infinite UV cut-off parameter.
Therefore, it might be unclear whether the higher-dimensional terms are generated when the UV cut-off is finite in the Fujikawa method, and we would like to derive 
the absence of the higher-dimensional gluon operators via the explicit one-loop calculation
without the introduction of the UV cut-off parameter.
Here, by using the operator-Schwinger method we show a theorem that the $T$-odd and $P$-odd higher-dimensional gluon operators are not generated 
at one-loop level even if the quark mass phases are not vanishing, whose diagram is shown in Fig.~\ref{fig:higher}.
\begin{figure}
    \centering
    \includegraphics[width=0.5\linewidth]{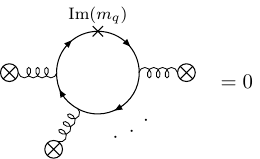}
    \caption{
        The one-loop babble diagram
        contributing to the radiative corrections to the $T$-odd and $P$-odd higher-dimensional gluon operators. It is shown that they are not induced from the quark mass phases. 
        }
    \label{fig:higher}
\end{figure}

In the operator-Schwinger method, 
we start from
the derivative of $\Delta S$ over $m_q$ for $P_R$ in Eq.~\eqref{dsdm}.
Then, by using the covariant derivative expansion (CDE)  \cite{Henning:2014wua,Henning:2016lyp}, 
the trace over the functional space can be evaluated by
\beq
&\frac{d}{d m_q}\Delta S
=
i \int d^4 x \int \frac{d^4 p}{(2 \pi)^4} 
\operatorname{tr} \left[
\frac{1}{p^2-|m_q|^2-(2 ip^\mu D_\mu + D^2+\frac12 g_s 
(\sigma \cdot G) ) } m_q^\star P_R
\right]
\\
& = i \int d^4 x \int \frac{d^4 p}{(2 \pi)^4} 
\operatorname{tr} \left\{
\left[\frac{1}{p^2-|m_q|^2}\left(\frac{1}{2} g_s (\sigma \cdot G)\right) \frac{1}{p^2-|m_q|^2}\left(\frac{1}{2} g_s (\sigma \cdot G) \right) \frac{1}{p^2-|m_q|^2} m_q^* P_R\right] \right.
\nonumber \\
&\quad + 
\operatorname{tr}  \left[\frac{1}{p^2-|m_q|^2}\left(\frac{1}{2} g_s (\sigma \cdot G)\right) \frac{1}{p^2-|m_q|^2}\left(\frac{1}{2} g_s (\sigma \cdot G)\right) \frac{1}{p^2-|m_q|^2}
\left(\frac{1}{2} g_s (\sigma \cdot G)\right) \frac{1}{p^2-|m_q|^2}
m_q^* P_R\right] \nonumber \\
& 
\quad +\mbox{(terms including two $G$s and two $D$s in the numerator)} \Biggr\}\nonumber \\
&  \quad + \cdots \,,
\label{eq:CDEexact}
\eeq
where the first line in Eq.~\eqref{eq:CDEexact} induces  $G\tilde{G}$ term, while the remaining ones we obtain 
\beq
\frac{d}{d m_q}\Delta S
&=
\int d^4 x \left(  \frac{i}{64 \pi^2} \frac{g_s^2}{m_q} G_{\mu \nu}^a \tilde{G}^{a \mu \nu} 
- \frac{i}{192\pi^2} r_{GGG} \frac{g_s^3}{m_q|m_q|^2 } f^{abc} G_{\mu \nu}^a G^{b \nu}_{\rho}\tilde{G}^{c \rho \mu}\right)\,,
\label{eq:delmq}
\eeq
where $r_{GGG}$ is a $\mathcal{O}(1)$ real number (for example, $r_{GGG}=1$ comes from the second line of Eq.~\eqref{eq:CDEexact}).
In the above equation, $G\tilde{G}$ term comes from 
the second order of  $\frac12 g_s 
(\sigma \cdot G)$ in the denominator and its coefficient is proportional to $1/m_q$. 
On the other hand, 
$G^2\tilde{G}$  term comes from 
the expansions of the covariant derivatives and/or $\frac12 g_s (\sigma \cdot G)$ and its coefficient is proportional to  $1/(m_q|m_q|^2)$. 
The coefficients are determined with the dimensions of the operators up to the $O(1)$ coefficients. 
Similarly, the derivative of $\Delta S$ over $m_q^\ast$ for $P_L$ is 
\beq
\frac{d}{d m_q^\ast}\Delta S
&=
i\operatorname{Tr}\left[
\frac{1}{P^2-|m_q|^2-\frac12 g_s\sigma G} m_q P_L
\right]
\nonumber \\
& = - \int d^4 x \left(  \frac{i}{64 \pi^2} \frac{g_s^2}{m_q^\ast} G_{\mu \nu}^a \tilde{G}^{a \mu \nu} 
- \frac{i}{192\pi^2}  r_{GGG} \frac{g_s^3}{m_q^\ast |m_q|^2 } f^{abc} G_{\mu \nu}^a G^{b \nu}_{\rho}\tilde{G}^{c \rho \mu}\right)\,.
\eeq
Therefore, the overall factor $1/m_q$ in Eq.~\eqref{eq:delmq} is replaced by  $-1/m_q^\star$, where the relative phase comes from the difference of the projection operators $P_{L/R}$.

The one-loop corrections to the effective action can be determined  
by adding the integrals $d\Delta S/d m_q$ with $m_q$ and $d\Delta S/d m_q^\star$ with $m_q^\star$, 
\beq
\Delta S & \supset \int d^4 x  \left( \frac{i g_s^2}{64\pi^2}  \log m_q G_{\mu \nu}^a \tilde{G}^{a \mu \nu} 
+ \frac{i g_s^3}{192 \pi^2}   r_{GGG} \frac{1}{|m_q|^2 }  f^{abc} G_{\mu \nu}^a G^{b \nu}_{\rho}\tilde{G}^{c \rho \mu}\right)\nonumber \\
& \quad + 
\int d^4 x  \left( -\frac{i g_s^2}{64\pi^2}  \log m_q^\ast G_{\mu \nu}^a \tilde{G}^{a \mu \nu} 
- \frac{i g_s^3}{192 \pi^2}   r_{GGG} \frac{1}{|m_q|^2 }  f^{abc} G_{\mu \nu}^a G^{b \nu}_{\rho}\tilde{G}^{c \rho \mu}\right)\nonumber \\
& =  \int d^4 x   \frac{i g_s^2}{64\pi^2}  \log \frac{m_q}{m_q^\ast} G_{\mu \nu}^a \tilde{G}^{a \mu \nu} \nonumber \\
& = -\int d^4 x  \theta_q \frac{  \alpha_s}{8\pi} G_{\mu \nu}^a \tilde{G}^{a \mu \nu}\,.
\eeq
Therefore, we find that only the QCD $\theta$ term is generated with the coefficient proportional to the quark mass phase, and the Weinberg operator is not induced from Fig.~\ref{fig:higher}.

We conclude that such a cancellation occurs for all the $T$-odd and $P$-odd higher-dimensional gluon operators with dimensions ($D$) larger than four.
In general,  $d\Delta S/d m_q$ includes all the $T$-odd and $P$-odd dimension-$D$ gluon operators
from the expansions of the covariant derivatives and/or $\frac12 g_s (\sigma \cdot G)$ in the denominator, and 
their coefficients are proportional to  $m_q^\star/|m_q|^{(D-2)}$. 
On the other hand, $d\Delta S/d m_q^\ast$ has the gluon operators 
with the coefficients proportional to $-m_q/|m_q|^{(D-2)}$. 
 By integrating $d\Delta S/d m_q$ with $m_q$ and integrating $d\Delta S/d m_q^\star$ with 
 $m_q^\star$,
 it is found that the coefficients of 
 the $T$-odd and $P$-odd higher-dimensional gluon operators in Fig.~\ref{fig:higher} are canceled, while only $D=4$ term survives. 
 Thus, we showed that the quark mass phases do not generate the $T$-odd and $P$-odd higher-dimensional gluon operators at the one-loop level.

\section{Two-loop contribution to QCD \texorpdfstring{\boldmath{$\theta$}}{theta}  parameter in simplified models}
\label{sec:2loop}

In this section, using the Fock-Schwinger gauge method, we evaluate the two-loop corrections to the QCD $\theta$ parameter in simple models, in which fermion(s) and a real scalar have $CP$-violating Yukawa interaction(s). We find that 
when the fermions and the scalar have masses comparable to each other, the Fujikawa method does not cover full contribution
in the evaluation of the QCD $\theta$ parameter even if taking into account the radiative correction to the imaginary parts of the fermion masses at one-loop level. 

\subsection{One flavor case}
First, 
we evaluate the two-loop contributions to the QCD $\theta$ parameter in a simplified model in which a $CP$-violating Yukawa coupling $y_{q}$ with colored fermion $q$ is contained, where the number of the flavor is one for simplicity. 
Let us consider the following Lagrangian,
\beq
-\mathcal{L} =  \frac{1}{2}m_\phi^2 \phi^2  + \bar{q} \left[ \text{Re}\left(m_{q}\right) + \text{Im}\left(m_{q}\right) i \gamma_5 \right]q +  y_q \bar{q} P_R q \phi + \text{h.c.}\,,
\eeq
where $\phi$ is a neutral scalar. 
Note that the bare QCD $\theta$ parameter $\theta_G$ is set to be zero, cf.\,Eq.~\eqref{eq:thetaG}.
In the following, we assume $ \text{Im}(m_{q}) \ll \text{Re}(m_{q})$ at the tree level in order to investigate the structure of the two-loop corrections using the low-energy effective theories. On the other hand, we keep the one-loop correction formula since the one-loop diagrams include the counter terms to cancel divergence of subdiagrams in the two-loop diagrams and the combined results of one- and two-loop corrections are renormalization-scale independent. 

\begin{figure}
    \centering
    \includegraphics[width=1\linewidth]{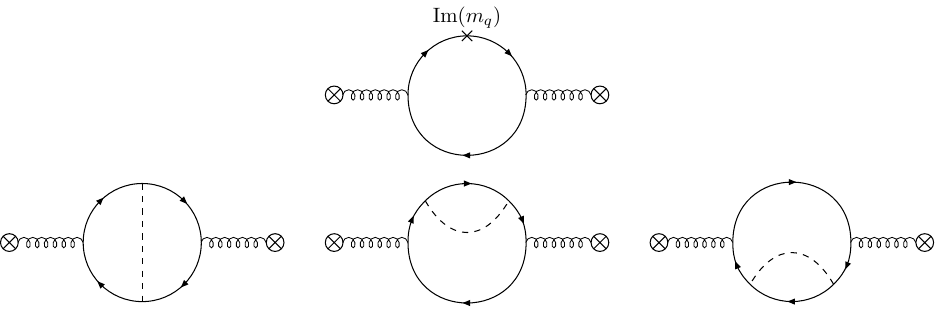}
    \caption{
        The one-loop and two-loop babble diagrams 
        which contribute to the radiative corrections to the QCD $\theta$ parameter.
        The one-loop diagram gives a simple result $\delta \theta|_{\rm 1L} = - \text{Im}(m_q)/\text{Re}(m_q)$. 
        For the two-loop diagrams, the first one generates $I_{(2;2)}$ loop function in Eq.~\eqref{eq:fullformula_theta}, while the others  
        $2 \bar{I}_{(3;1)}$ in total.
        }
    \label{fig:babbles}
\end{figure}
When one sets $\theta_G=0$, the physical QCD $\theta$ parameter can be perturbatively calculated 
by the Fock-Schwinger gauge method \cite{Hisano:2023izx,Novikov:1983gd}.
The radiative corrections $\delta \theta$ are derived from one-loop (1L) and two-loop diagrams (2L) given in Fig.~\ref{fig:babbles}, $\delta \theta = \delta\theta|_{\rm 1L}
+ \delta\theta|_{\rm 2L}$, as \cite{Hisano:2023izx}
\beq
\bar{\theta} &= \delta \theta = \delta\theta|_{\rm 1L}  + \delta\theta|_{\rm 2L} + \cdots \,, \\
\delta\theta|_{\rm 1L}&=-
\frac{\text{Im}(m_q)}{\text{Re}(m_q)}\,, 
\label{eq:1L}
\\
  \delta\theta|_{\rm 2L}&=\frac{{\rm Im}\left(y_q^2  \right)}{16\pi^2} {\left[\text{Re}(m_q)\right]^2 }
  \left[I_{(2;2)}(m_q^2; m_q^2;m_\phi^2)+2 \bar{I}_{(3;1)}(m_q^2; m_q^2; m_\phi^2)\right]\,,
\label{eq:fullformula_theta}
\eeq
with\footnote{%
In the model, $\theta_G-\theta_q$ ($m_q=|m_q|\exp(i \theta_q)$) and $y_q m_q^\ast$ are invariant under the  chiral rotation of the colored fermion. In addition, the Lagrangian is invariant under $\phi\rightarrow -\phi$ and $y_q \rightarrow -y_q$. These imply that  
an accurate numerator of the two-loop correction is
$\text{Im}((y_q m_q^\ast)^2)$, 
and we expand it with $ \text{Im}(m_{q}) \ll \text{Re}(m_{q})$.
}
\beq
 \bar{I}_{(3;1)}(x_1;x_2;x_3)&=  I_{(3;1)}(x_1;x_2;x_3)
-\frac{1}{2 x_1} \frac{1}{\bar\epsilon} \left(1 -\epsilon \log\frac{x_1}{Q^2}\right)\nonumber \\
& = I_{(3;1)}(x_1;x_2;x_3)
-\frac{1}{2 x_1} \left( \frac{1}{\bar\epsilon}
- \log\frac{x_1}{Q^2}\right)\,,
\label{eq:ibar31}
\eeq
where $Q^2\equiv4\pi \mu^2 {\rm e}^{-\gamma_E}$ and
the dimensional regularization 
is used with 
the renormalization scale.
Here, 
the mass squared for the colored-quark masses in the loop functions is understood as 
$m_i^2 =  {\text{Re}(m_i)}^2+{\text{Im}(m_i)}^2$.
The two-loop functions  $I_{(2;2)}$ and $I_{(3;1)}$ are given in Appendix~\ref{app:loop}. 

The loop-function $I_{(3;1)}$ corresponds to the last two two-loop diagrams in Fig.~\ref{fig:babbles}, whose subdiagram gives a one-loop UV divergence $\propto y_q^2 m_q /\bar{\epsilon}$, and this divergence must be canceled by a counter term of the mass renormalization.
We adopt the $\overline{\rm MS}$ scheme for the mass renormalization.
The second term in $\bar{I}_{(3;1)}(x_1;x_2;x_3)$, subtracting a $1/\bar{\epsilon}$ pole, corresponds to the counter term contribution,
and the factor $(1 -\epsilon \log x_1/Q^2)$ comes from the higher-order term of $\epsilon$ in the dimensional regularization (see Eq.~(2.15) in Ref.~\cite{Hisano:2023izx}).\footnote{%
The second term in $\bar{I}_{(3;1)}(x_1;x_2;x_3)$ corresponds to 
the divergence contribution $\bar{I}_{\epsilon(3;1)}$ defined in  Ref.~\cite{Hisano:2023izx} when one adopts 
the ${\rm MS}$ scheme for the mass renormalization. As a result,
$Q^2$ in ${F}_0(p^2,x_1,x_2)$ in Ref.~\cite{Hisano:2023izx} is replaced with
$\mu^2$ in $F_0(p^2,x_1,x_2)$ in Eq.~(\ref{func_f}). 
}
Then, $\bar{I}_{(3;1)}(x_1,x_2,x_3)$ is UV and IR finite function \cite{Hisano:2023izx}.

The above formula for the radiative corrections to the QCD $\theta$ parameter at two-loop level becomes simpler when the colored fermion is much lighter (or heavier) than the scalar $\phi$, since the effective theory description works. 
Let us check such hierarchical situations. 

First, when the scalar particle mass is much heavier than the fermion mass $m_\phi \gg m_q$, 
the two-loop contributions in Eq.~\eqref{eq:fullformula_theta} can be simplified to 
\beq
& {I}_{(2;2)}(m_q^2; m_q^2;m_\phi^2) + 2\bar{I}_{(3;1)}(m_q^2; m_q^2;m_\phi^2)
\nonumber\\
 &\quad\quad \simeq 
 \frac{1}{ m_q^2}\left(1-\log\frac{m_\phi^2}{\mu^2}\right)
+\frac{\pi^2+ 12 \log\frac{m_q^2}{m_\phi^2}
  +3\log^2\frac{m_q^2}{m_\phi^2}}{3 m_\phi^2}\,.
\label{eq:largex3}  
\eeq

From the viewpoint of the effective field theory, first of all, the scalar $\phi$ is integrated out at a scale of $\mu=m_\phi$, then the following $CP$-violating effective interactions may be obtained up to the dimension-six operators \cite{Hisano:2012cc},
\beq
-\mathcal{L}_{\rm eff}& =
\bar{q} \left[ 
\text{Im}\left(m_{q}+ \Delta m_q \right) i \gamma_5 \right]q 
\nonumber \\
& \quad + \frac{i}2 g_s \tilde{d}_q  \bar{q}  \sigma^{\mu\nu} \gamma_5 T^a q G_{\mu\nu}^a 
- C_4^q (\bar{q}q)(\bar{q} i \gamma_5 q) 
- C_5^q(\bar{q} \sigma^{\mu\nu}q)(\bar{q} i \sigma_{\mu \nu}\gamma_5 q) 
\nonumber \\
& \quad + \Delta \theta|_{\rm th} \frac{\alpha_s}{8\pi}G^a_{\mu \nu} \tilde G^{a \mu \nu} + \frac{1}{3}\omega f^{abc} G^{a}_{\mu \nu} G^{b \nu}_{\rho}\tilde{G}^{c \rho \mu}\,, 
\label{eq:1fLeff}
\eeq
where $f^{abc}$  is the structure constant. 

The radiative correction to the imaginary part of the fermion mass at $\mu \simeq m_\phi$ is given as 
\beq
  \text{Im} \left[ \Delta m_{q}(\mu)\right]&=
 \frac{{\rm Im}\left(y_{q}^2\right)}{16\pi^2} {\text{Re}(m_q)}
 {F}_0(0,m_q^2,m^2_\phi) \,,
 \label{zeromomentummass}
 \eeq
where
\beq
{F}_0(p^2,x_1,x_2)
&=
\int_0^1 dz \log\frac{-z(1-z) p^2+z x_1+(1-z) x_2}{\mu^2}
\label{func_f}\,.
\eeq
The above radiative correction to the imaginary part of the fermion mass corresponds to that in a limit of the zero external momentum, since the external momentum is negligible compared with $m_\phi$. In the case, the radiative correction is explicitly given as 
\beq
\text{Im} \left[ \Delta m_{q}(\mu)\right]
 &=  -\frac{{\rm Im}\left(y_{q}^2\right)}{16\pi^2} {\text{Re}(m_q)}
  \left[ 1 - \frac{1}{m_q^2-m_\phi^2 }
  \left(m_q^2\log\frac{m_q^2}{\mu^2}-m_\phi^2\log\frac{m_\phi^2}{\mu^2}\right)\right]\\
&\simeq 
\left\{
\begin{array}{ll}
\frac{{\rm Im}\left(y_{q}^2\right)}{16\pi^2} {\text{Re}(m_q)}
  \log\frac{m_\phi^2}{\mu^2}
  & \quad \text{for~} m_\phi \approx m_q \,,\\
-\frac{{\rm Im}\left(y_{q}^2\right)}{16\pi^2} {\text{Re}(m_q)}
  \left(1-\log\frac{m_\phi^2}{\mu^2}
  +
  \frac{m_q^2}{m_\phi^2} \log \frac{m_q^2}{m_\phi^2}\right)
   & \quad \text{for~} m_\phi \gg m_q \,.
   \label{eq:1f loop mass phase}
\end{array}
\right.  
\eeq

It is found by explicit calculation that the matching conditions for the higher-dimensional operators in Eq.~(\ref{eq:1fLeff}) at $\mu = m_\phi$ are\footnote{
The chiral rotation of $q$ to remove ${\rm Im}[\Delta m_q]$ leads to the contribution to the CEDM from the CMDM in Eq.~\eqref{eq:physicaldq}. However, since ${\rm Im}[\Delta m_q]$ is at one-loop level, the contribution to the CEDM is at two-loop level, and then we ignore it.}
}
\beq
\tilde{d}_q &= 
  -\frac{{\rm Im}\left(y_{q}^2\right)}{16\pi^2} \frac{3{\text{Re}(m_q)}}{2m_\phi^2}\,, \\
C_4^q & =  \frac{\text{Im}\left(y_{q}^2\right)}{2 m_\phi^2}\,,\\
C_5^q &= \omega = 0\,.
\label{wilsoncoefficients}
\eeq
The  term $\Delta \theta|_{\rm th}$ in Eq.~\eqref{eq:1fLeff} 
is a threshold correction to the QCD $\theta$ parameter, which will be identified by comparing with the full result and the contributions from the effective operators in the effective theory.

In the effective Lagrangian \eqref{eq:1fLeff},
the CEDM operators are mixed with the four-Fermi operators via the renormalization-group equation \cite{Hisano:2012cc}, and then the $\tilde{d}_q$ depends on $\mu$ when $m_q<\mu<m_\phi$. Thus,  $\tilde{d}_q$ is approximately given as 
\begin{eqnarray}
  \tilde{d}_q(\mu) &=&
  -\frac{{\rm Im}\left(y_{q}^2\right)}{16\pi^2} \frac{{\text{Re}(m_q)}}{m_\phi^2}
  \left(\frac{3}{2}+\log\frac{\mu^2}{m_\phi^2}\right)\,.
  \label{eq:1f cedm}
\end{eqnarray}
This CEDM operator induces the QCD $\theta$ parameter via the renormalization-group equation \cite{Jenkins:2017dyc}, as reviewed in Sec.~\ref{sec:1loop}.

Using these results in the effective theory, 
the radiatively generated QCD $\theta$ parameter at the low-energy scale is evaluated as 
\beq
\delta \theta 
&=
-\frac{\text{Im} (m_{q}+\Delta m_{q})}{\text{Re} (m_{q})}
 -2 \int^{\log m^2_q}_{\log m^2_\phi} \tilde{d}_q(\mu) {\text{Re} (m_{q})}  d(\log\mu^2)
+\left. \Delta \theta\right|_{\rm th}\\
& = 
-\frac{\text{Im} (m_{q})}{\text{Re} (m_{q})} 
+\frac{{\rm Im}\left(y_{q}^2\right)}{16\pi^2} 
  \left(1-\log\frac{m_\phi^2}{\mu^2}
  +
  \frac{m_q^2}{m_\phi^2} \log \frac{m_q^2}{m_\phi^2}\right)
\nonumber \\
& \quad + \frac{{\rm Im}\left(y_{q}^2\right)}{16\pi^2} \frac{m_q^2}{m_\phi^2}\left( {3} \log \frac{m_q^2}{m_\phi^2} +  \log^2 \frac{m_q^2}{m_\phi^2}  \right)
+ \left. \Delta \theta\right|_{\rm th}\,.
\eeq
This result is consistent with the QCD $\theta$ parameter directly calculated from the diagrams in Fig.~\ref{fig:babbles} (see Eq.~\eqref{eq:largex3}),
\beq
\delta \theta 
&= 
\left.\delta \theta\right|_{1L}+\left.\delta \theta\right|_{2L}\nonumber\\
& = - \frac{\text{Im}(m_q)}{\text{Re}(m_q)}
+ \frac{{\rm Im}\left(y_{q}^2\right)}{16\pi^2}
 \left[1-\log\frac{m_\phi^2}{\mu^2}
+\frac{m_q^2}{3 m_\phi^2 } \left(\pi^2+ 12 \log\frac{m_q^2}{m_\phi^2}
  +3\log^2\frac{m_q^2}{m_\phi^2}\right)\right]\,,
\eeq
where $\mu$ should be taken for the heaviest particle mass of the loop diagrams, $\mu \simeq m_\phi$.
In addition, by matching the above results one obtains the threshold correction to the QCD $\theta$ parameter as 
\beq
\left. \Delta \theta\right|_{\rm th}
=\frac{{\rm Im}\left(y_{q}^2\right)}{48} \frac{{m_{q}^2}}{m_\phi^2}\,.
\label{eq:1flavor threshold}
\eeq
Note that these effective theory matchings are valid only when $m_\phi \gg 2 m_q$ \cite{Davydychev:1992mt}.

In the above calculation, we implicitly assume that the light fermion is heavier than the $\Lambda_{\rm QCD}$ scale. On the other hand, the light quarks in the SM cannot be integrated out. 
The neutron EDM is evaluated in the low-energy effective theory, which includes higher-dimensional operators of light quarks in addition to those of gluons. 
Note that when fermions with mass larger than $\Lambda_{\rm QCD}$, such as heavier quarks in the SM, 
are integrated out, the fermion CEDMs induce the Weinberg operator as a threshold correction and it also contributes to the neutron EDM \cite{Boyd:1990bx,Braaten:1990gq,Chang:1990jv,Dine:1990pf,Abe:2017sam},
though it does not contribute to the QCD $\theta$ term. 
The contribution to the neutron EDM 
from the induced Weinberg operator would be smaller than the contribution from $\delta \theta$, suppressed by $\mathcal{O}(\Lambda_{\rm QCD}^2/m_\phi^2)$.

\begin{figure}[t]
\begin{center}
\begin{subfigure}{0.48 \textwidth}
\includegraphics[width = \textwidth]{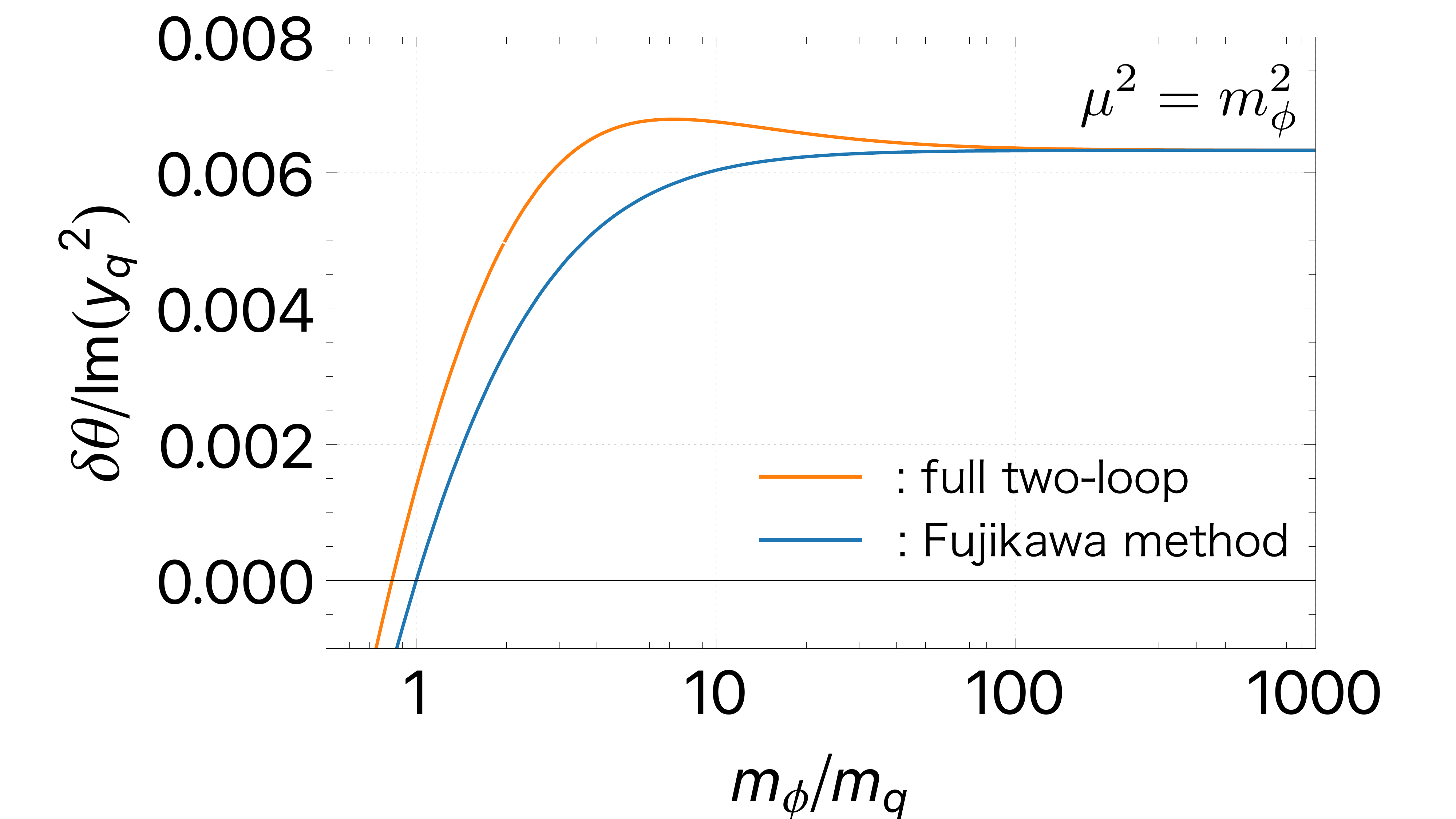}
\caption{$m_\phi \gsim m_q$}
\label{fig:1flavor left}
\end{subfigure}
\quad 
\begin{subfigure}{0.48 \textwidth}
\includegraphics[width = \textwidth]{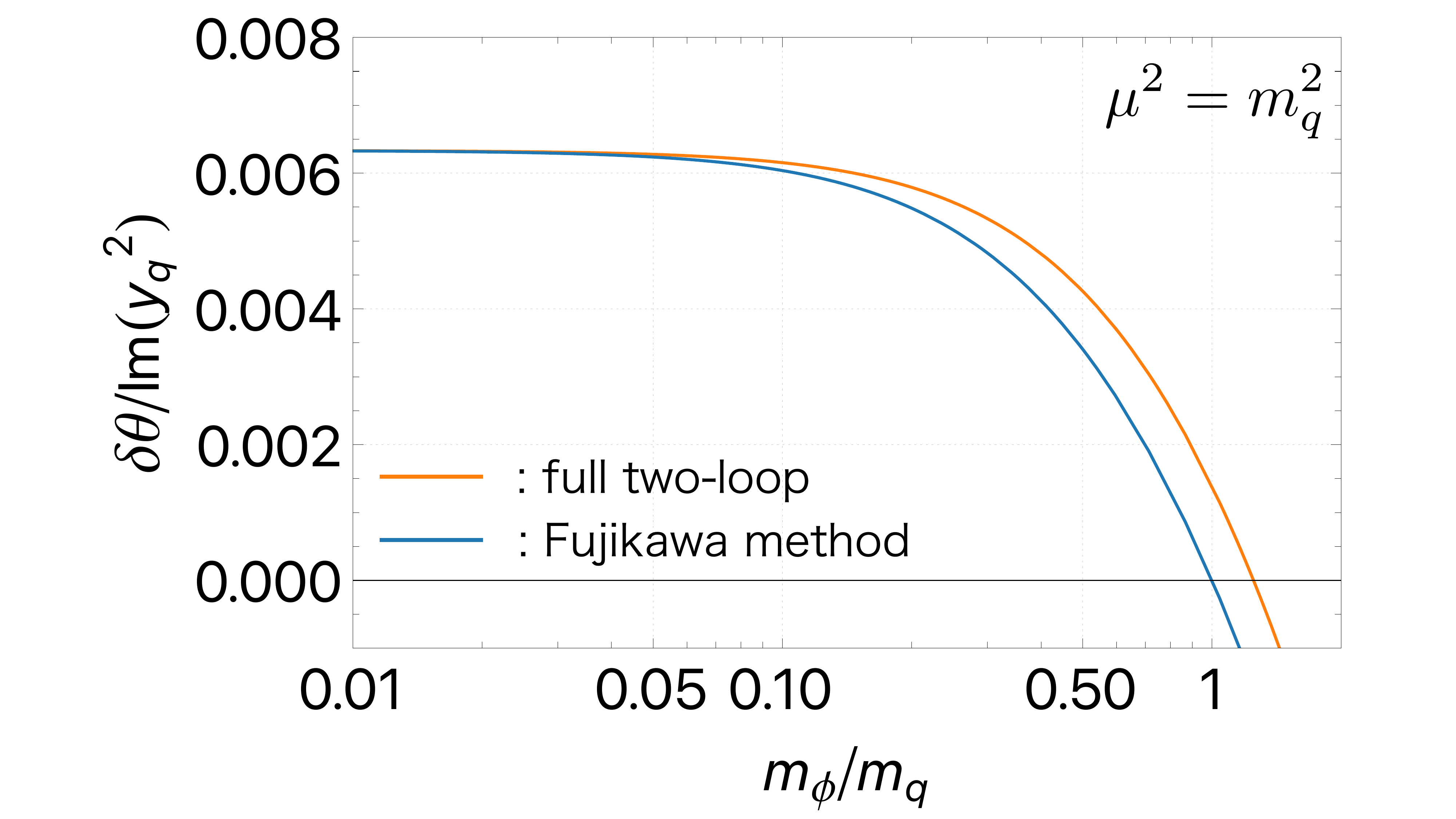}
\caption{$m_\phi \lsim m_q$}
\label{fig:1flavor right}
\end{subfigure}
\caption{
The radiative corrections to the QCD $\theta$ parameter, $\delta \theta$ which is normalized by $\textrm{Im}(y_q^2)$, evaluated from the full two-loop diagrams (Eq.~(\ref{eq:fullformula_theta})) are presented by the orange lines.
For comparison, the corrections evaluated 
from the one-loop correction to the colored fermion mass in a limit of zero external momentum (Eq.~(\ref{zeromomentummass})) with the Fujikawa method are also shown by the blue lines. 
In the left panel (a),
a parameter region of 
$m_\phi \gtrsim m_q$ is shown with 
$\mu=m_\phi$. On the other hand, 
in the right panel (b) $m_\phi \lesssim m_q$ with $\mu=m_q$. 
}
\label{fig:1flavor}
\end{center}
\end{figure}

We show the numerical examination in Fig.~(\ref{fig:1flavor left}) in order to stress the importance of the full two-loop diagram calculation in the case of  $m_\phi \gtrsim m_q$. Here, $\mu=m_\phi$, since the effective theory of the colored fermion works after integrating out $\phi$, as mentioned above.
The blue line represents the contribution to the QCD $\theta$ parameter from the imaginary part of the colored fermion mass in Eq.~(\ref{zeromomentummass}) with the Fujikawa method in Eq.~\eqref{eq:argdetm}, while the orange one comes from the direct two-loop diagram calculation in Eq.~(\ref{eq:fullformula_theta}).
The CEDM of the colored fermion contributes to the QCD $\theta$ parameter, though it is suppressed by $m_q^2/m^2_\phi$. 
The threshold correction in Eq.~\eqref{eq:1flavor threshold} is also 
suppressed by $m_q^2/m^2_\phi$. 
As a result, the correction to the QCD $\theta$ parameter is dominated by the contribution from the correction to the imaginary part of the colored fermion mass when $m_\phi \gg m_q$, as expected in Ref.~\cite{Khriplovich:1993pf}.
However, it is also shown that when $m_\phi/m_q \lesssim 10$, the estimation of the Fujikawa method does not cover full contribution.\\

On the other hand, when the scalar particle mass is quite light compared to the fermion mass $m_\phi \ll  m_q$, 
the two-loop functions can be simplified to \cite{Davydychev:1992mt}
\beq
& {I}_{(2;2)}(m_q^2; m_q^2;m_\phi^2) + 2\bar{I}_{(3;1)}(m_q^2; m_q^2;m_\phi^2)\nonumber\\
 &\quad\quad 
 \simeq 
\frac{1}{m_q^2}\left( 1 - \log \frac{m_q^2}{\mu^2}\right)\,,
\eeq
where all two-loop diagrams are contributing.\footnote{%
\label{footnote2loop}
For  $m_\phi \ll  m_q$, each two-loop function becomes
\beq
{I}_{(2;2)}(m_q^2; m_q^2;m_\phi^2) &= \frac{1}{m_q^2} + \mathcal{O}\left(\frac{m_\phi^2}{m_q^4}\right)\,,\\
{I}_{(3;1)}(m_q^2; m_q^2;m_\phi^2)  &= - \frac{1}{2 m_q^2} \log \frac{m_q^2}{\mu^2} + \mathcal{O}\left(\frac{m_\phi^2}{m_q^4}\right)\,.
\eeq
When one sets $\mu=m_q$, the two-loop contribution is predominated by ${I}_{(2;2)}(m_q^2; m_q^2;m_\phi^2) $.
}

From the viewpoint of the effective theory, this contribution corresponds to a threshold correction to the QCD $\theta$ parameter when the fermion is integrated out,
\beq
\left. \Delta \theta\right|_{\rm th} = -\frac{{\rm Im}(m_q)}{{\rm Re}(m_q)}+
 \frac{\text{Im}\left(y_q^2\right)}{16 \pi^2} \left(1 - \log \frac{m_q^2}{\mu^2}\right)\,,
\eeq
and hence $\mu \simeq m_q$. 
The $\mu$ dependence is canceled between the first and the second terms.
It is found that the second term coincides with the one-loop  correction to the imaginary part of the fermion mass (${\rm Im}[\Delta m_q(\mu)]$) in a limit of the zero external momentum, normalized by the fermion mass, 
\beq
- \frac{{\rm Im}[\Delta m_q(\mu)]}{{\rm Re}(m_q)}
&=-  \frac{{\rm Im}(y_q^2)}{16\pi^2}F_0(0,m_q^2,0)
= \frac{\text{Im}\left(y_q^2\right)}{16 \pi^2} \left(1 - \log \frac{m_q^2}{\mu^2}\right)\,.
\eeq

In Fig.~(\ref{fig:1flavor right}), we take $m_\phi \lesssim m_q$. Here, $\mu=m_q$.
Again, the blue line represents the contribution to the QCD $\theta$ parameter from the imaginary part of the colored fermion mass in Eq.~(\ref{zeromomentummass}) 
with the Fujikawa method in Eq.~\eqref{eq:argdetm}, while the orange one comes from the direct two-loop diagram calculation in Eq.~(\ref{eq:fullformula_theta}). 
The blue and orange lines  coincide when $m_\phi \ll m_q$, 
while all three diagrams at two-loop level contribute to the QCD $\theta$ parameter (see footnote\ref{footnote2loop}), as mentioned above.

Thus, when $m_\phi \gg m_q$ or $m_\phi\ll m_q$, 
the radiative correction to the QCD $\theta$ parameter at two-loop level can be sufficiently evaluated by the Fujikawa method taking into account the radiative correction to the fermion masses in a limit of the zero external momentum. 
The coincidence is, however, non-trivial for the case of $m_\phi\ll m_q$. 
It would be an open question whether the coincidence is valid even if one includes the higher-order corrections to the QCD $\theta$ parameter than the two-loop one.

\subsection{Two hierarchical flavor case}

Next, we consider a more complicated situation:
there are two kinds of fermions $q_l$ (light  fermion, $m_l \ll m_\phi$) and $q_h$ (heavy one, $m_h \approx m_\phi$), and the interactions are
\beq
-\mathcal{L} = \bar{q}_i \left[ \text{Re}\left(m_{i}\right) + \text{Im}\left(m_{i}\right) i \gamma_5 \right]q_i + \left( \bar{q}_i y_{ij} P_R q_j \phi + \text{h.c.}\right)\,,
\eeq
where $i,j$ run $l$ (light) and $h$ (heavy), and we take a 
flavor diagonal mass basis.
We suppose that the Yukawa interactions $y_{lh}$ and $y_{hl}$  are $CP$-violating complex couplings but $y_{ll}$ and  $y_{hh}$ are real ones, for simplicity.

From the direct loop calculations, 
the radiative corrections to the QCD $\theta$ parameter are
\beq
\delta\theta|_{\rm 1L}&=-
\sum_i \frac{\text{Im}(m_i)}{\text{Re}(m_i)} \nonumber \\
& = 
-
 \frac{\text{Im}(m_l)}{\text{Re}(m_l)}
 -
\frac{\text{Im}(m_h)}{\text{Re}(m_h)}
\,, 
\label{eq:1L2f}
\\
  \delta\theta|_{\rm 2L}&=\sum_{i,j}\frac{{\rm Im}\left(y_{ij}y_{ji}\right)}{16\pi^2} {\text{Re}(m_i)}{\text{Re}(m_j)}
  \left[I_{(2;2)}(m_i^2; m_j^2;m_\phi^2)+2\bar{I}_{(3;1)}(m_i^2; m_j^2; m_\phi^2)\right]\nonumber \\
&=  \frac{{\rm Im}\left(y_{lh}y_{hl}\right)}{8\pi^2} {\text{Re}(m_l)}{\text{Re}(m_h)}\nonumber\\
&\times
  \left[  I_{(2;2)}(m_l^2; m_h^2;m_\phi^2)
  +\bar{I}_{(3;1)}(m_l^2; m_h^2; m_\phi^2)+\bar{I}_{(3;1)}(m_h^2; m_l^2; m_\phi^2)\right]
  \,.
\label{eq:fullformula_theta2f}
\eeq
Here, $ I_{(2;2)}(x_1;x_2;x_3)= I_{(2;2)}(x_2;x_1;x_3)$ is used.
When $m_l \ll m_h,\,m_\phi$, these loop functions can be simplified to
\beq
&  {I}_{(2;2)}(m_l^2;m_h^2;m_\phi^2)+ \bar{I}_{(3;1)}(m_l^2;m_h^2;m_\phi^2)+\bar{I}_{(3;1)}(m_h^2;m_l^2;m_\phi^2)
\nonumber\\
 &\quad\quad =\frac{1}{2m_l^2}\left[1-\frac{1}{m_h^2-m_\phi^2}\left(m_h^2\log\frac{m_h^2}{{\mu^2}}-m_\phi^2\log\frac{m_\phi^2}{\mu^2}\right)\right]\nonumber\\
&\quad\qquad +\frac{1}{2m_h^2}\left[1-\frac{1}{m_h^2-m_\phi^2}\left(m_h^2\log\frac{m_h^2}{\mu^2}-m_\phi^2\log\frac{m_\phi^2}{\mu^2}\right)\right]\nonumber\\
&\quad\qquad -\frac{m_h^4-4m_\phi^2 m_h^2+3m_\phi^4+2 m_\phi^4 \log\frac{m_h^2}{m_\phi^2}}{2(m_h^2-m_\phi^2)^3}\log\frac{m_l^2}{m_\phi^2}\nonumber\\
&\quad\qquad +\frac{(m_h^4- 3m_h^2 m_\phi^2 -2m_\phi^4 ) \log\frac{m_h^2}{m_\phi^2}
-4m_\phi^4~{\rm Li}_2\left(1-\frac{m_h^2}{m_\phi^2}\right)}{2(m_h^2-m_\phi^2)^3}\,.
\label{eq:2ffullapp}
\eeq

Let us compare the radiative corrections to the QCD 
$\theta$ parameter in the effective theory approach. 
When the heavy fermion and the scalar boson are integrated out from the full theory, the following effective interactions are obtained,
\beq
-\mathcal{L}_{\rm eff}& =
\bar{q}_l \left[ \text{Re}\left(m_{l} + \Delta m_l \right) + \text{Im}\left(m_{l}+ \Delta m_l \right) i \gamma_5 \right]q_l
\nonumber \\
& \quad + \frac{i}2 g_s \tilde{d}_{q_l}  \bar{q}_l  \sigma^{\mu\nu} \gamma_5 T^a q_l G_{\mu\nu}^a 
+ \Delta \theta|_{\rm th} \frac{\alpha_s}{8\pi}G^a_{\mu \nu} \tilde G^{a \mu \nu}\,.
\label{eq:2fLeff}
\eeq
In addition to the Weinberg operator, 
the $CP$-violating four-Fermi operators are not generated 
since it is assumed that only $y_{lh}$ and $y_{hl}$ have complex phases.  The one-loop correction $\text{Im} (\Delta m_{l})$,
which is the correction in a limit of the zero external momentum,  is
\beq
  \text{Im} [\Delta m_{l}(\mu)]&=
\frac{{\rm Im}\left( y_{lh}y_{hl}\right)}{16\pi^2}  {\text{Re}(m_h)}F_0(0,m_h^2,m_\phi^2)\nonumber\\
&=
  -\frac{{\rm Im}\left( y_{lh}y_{hl}\right)}{16\pi^2}  {\text{Re}(m_h)}
  \left[ 1 - \frac{1}{m_h^2-m_\phi^2 }
  \left(m_h^2\log\frac{m_h^2}{\mu^2}-m_\phi^2\log\frac{m_\phi^2}{\mu^2}\right)\right]\,.
\label{eq:rc_quarkmass}
\eeq
Note that the one-loop corrected mass $(m_{l}+\Delta m_{l})$ is $\mu$-independent because $m_l$ is the $\overline{\rm MS}$ masses. 
In addition,  
the CEDM for $q_l$ at one-loop level is given as 
\begin{eqnarray}
  \tilde{d}_{q_l}&=&
  -\frac{{\rm Im}\left(y_{lh}y_{hl}\right)}{16\pi^2} \frac{{\text{Re}(m_h)}}{m_\phi^2}
  f\left(\frac{m_h^2}{m_\phi^2}\right)\,,
\end{eqnarray}
where 
\begin{eqnarray}
  f(x)&=&\frac{1}{2(1-x)^3}(3-4x+x^2+2\log x)\,.
\end{eqnarray}  

Using the above results, 
the radiative correction to the QCD $\theta$ parameter at two-loop level in Eq.~(\ref{eq:2ffullapp}) is reduced to 
\beq
\delta \theta &= \left.\delta \theta\right|_{1L}+\left.\delta \theta\right|_{2L}\nonumber\\
&=
-\frac{\text{Im} \left[m_{l}+\Delta m_{l}(\mu)\right]}{\text{Re} (m_{l})}
-2 \tilde{d}_{q_l} {\text{Re} (m_{l})}\log\frac{m_l^2}{m_\phi^2} +\left.\Delta \theta\right|_{\rm th}\,.
\eeq
The first two terms come from the integration of $q_l$ in the effective theory. 
The CEDM contribution is proportional to a large log, $\log (m_l/m_\phi)$.
It is expected since 
the renormalization-group equation of $\theta$ involves a term proportional to the CEDM in Eq.~\eqref{eq:RGE}.
These results are easily improved by the renormalization-group equations. 
The threshold correction
$\Delta \theta|_{\rm th}$
represents the contributions that come from the integration of $q_h$ and $m_\phi$ in the full theory, and we obtain it from the above matching as, 
\beq
\left.\Delta \theta\right|_{\rm th}
&=-\frac{\text{Im} \left[m_{h} +\Delta m_h(\mu)\right]}{\text{Re} (m_{h})}
\nonumber\\
&\quad 
+
\frac{{\rm Im}\left(y_{lh}y_{hl}\right)}{16\pi^2} {\text{Re} (m_{l})}{\text{Re} (m_{h})}
\frac{\left(
m^2_hm_\phi^2 + 3m^4_\phi
 \right)\log\frac{m^2_h}{m^2_\phi}
+ 4m^4_\phi~{\rm Li}_2\left(1-\frac{m^2_h}{m_\phi^2}\right)}{(m^2_\phi-m^2_h)^3}
\,.
\eeq
Here, $\text{Im} [\Delta m_{h}(\mu)]$ is the radiative correction to the imaginary part of heavy fermion mass in a limit of the zero external momentum,
\beq
  \text{Im} [\Delta m_{h}(\mu)]&=
 -\frac{{\rm Im}\left( y_{lh}y_{hl}\right)}{16\pi^2}  {\text{Re}(m_l)}
  \left(1-\log\frac{m_\phi^2}{\mu^2}
  \right)\,. 
\eeq

The threshold correction to the QCD $\theta$ parameter is suppressed by $\text{Re} (m_{l})/\text{Re} (m_{h})$, since we assume that 
$y_{lh}$ and $y_{hl}$  are $CP$-violating complex couplings while $y_{ll}$ and  $y_{hh}$ are real ones.
The CEDM contribution 
in the effective theory is also suppressed by 
$\text{Re} (m_{l})\text{Re} (m_{h})/m_\phi^2$. Thus, the correction to the QCD $\theta$ parameter can be evaluated by the 
Fujikawa method taking into account the radiative correction to the imaginary part of the light quark mass in Eq.~\eqref{eq:rc_quarkmass} in this case.

\section{Nelson-Barr Model}
\label{sec:NB}
In this section, we show an example in which the QCD $\theta$ parameter is induced at two-loop level while it vanishes at tree level. The $CP$ symmetry is spontaneously broken in the Nelson-Barr models \cite{Nelson:1983zb,Barr:1984qx}. The QCD $\theta$ term vanishes at the tree level by assuming quark mass matrices ${\cal M}_q(q=u,d)$ with  ${\rm arg\,det}\,({\cal M}_q)=0$ even after the spontaneous $CP$-violating sector.

The minimal models were derived by Bento, Branco, and Parada \cite{Bento:1991ez}. They are classified into two types of models, depending on whether up- or down-type quarks are coupled with the $CP$-violating sectors. In this paper, we consider the $d$-type model,\footnote{
The $u$-type model is given by 
\begin{eqnarray}
  -{\cal L}^u_Y&=&y_{u}^{ij}\tilde{H}\bar{Q}_i u_j+y_{d}^{ij}{H}\bar{Q}_i d_j+
  g^{aj}\Sigma_a \bar{\psi}u_j + m \bar{\psi}\psi^c +\textrm{h.c.}\,.
\nonumber
\end{eqnarray}
} given by
\begin{eqnarray}
  -{\cal L}^d_Y&=&y_{u}^{ij}{\tilde{H}}\bar{Q}_i u_j+y_{d}^{ij}{H}\bar{Q}_i d_j+
  g^{aj}\Sigma_a \bar{\psi} d_j + m\bar{\psi}\psi^c +\textrm{h.c.}\,,
\end{eqnarray}
where the down-type quarks are coupled with the $CP$-violating sector. Here,
$Q$, $u$, and $d$ are for $SU(2)_L$ doublet left-handed quarks, singlet right-handed up and down quarks, respectively. The subscripts $i,j,k$ are used for the generation in the SM ($i,j,k = 1,2,3$).
We introduce an $SU(2)_L$ singlet vector-like quark, whose left and right-handed components are represented by $\psi$ and $\psi^c$, respectively. The vacuum expectation value of the SM Higgs doublet field $H$ is $\langle H\rangle =(0,v/\sqrt{2})^T$.  The conjugate is $\tilde{H}=\epsilon H^\star$. The complex singlet scalar fields $\Sigma_a$ ($a=1,\cdots, N_\Sigma$, $N_\Sigma>1$) are assumed to get the complex vacuum expectation values so that the $CP$ symmetry is spontaneously broken. 
The gauge charges of the matter contents are summarised in Table~\ref{tab:matter contents}.
In this model, the down-type quark mass matrix is given by
\begin{eqnarray}
  \left(
  \begin{array}{cc}\bar{Q}&\bar{\psi}\end{array}\right){\cal M}
_d  \left(\begin{array}{c}
    d\\ \psi^c
  \end{array}\right)
  &=&
  \left(\begin{array} {cc}\bar{Q} & \bar{\psi}\end{array}\right)
  \left(\begin{array}{cc}
    y_d\langle {H}\rangle& 0\\
    \xi^\dagger& m
  \end{array}
    \right)
  \left(\begin{array}{c}
    d\\ \psi^c
  \end{array}\right)  \,,
\end{eqnarray}
where $\xi^\star_i=g^{ai}\langle \Sigma_a\rangle$.  We have to forbid terms such as $\Sigma_a \bar{\psi}\psi^c$ and ${H}\bar{Q}\psi^c$ since they lead to ${\rm arg\,det}({\cal M}_d)\ne 0$. Those terms are forbidden by  the following discrete $\mathbb{Z}_N$ symmetry 
imposed as,
\begin{eqnarray}
  \Sigma_a&\rightarrow& {\rm e}^{\frac{2\pi i k}{N}}\Sigma_a\,, \nonumber\\
  \psi&\rightarrow& {\rm e}^{\frac{2\pi i k}{N}}\psi\,, \\
  \psi^c&\rightarrow& {\rm e}^{\frac{2\pi i k}{N}}\psi^c\,. \nonumber
\end{eqnarray}

    %======================================================
    \begin{table}[t]
        \centering
        \newcommand{\bhline}[1]{\noalign{\hrule height #1}}
        \renewcommand{\arraystretch}{1.5}
        \rowcolors{2}{gray!15}{white}
        \begin{tabular}{ c c c c c c}
        \bhline{1 pt}
             &
             chirality & 
            $SU(3)_C$      &
            $SU(2)_L$      &
            $U(1)_Y$      &
            $\mathbb{Z}_N$   
            \\
                  \hline    
            $Q$      &
            L&  
            $\bf 3$      &
            $\bf 2$      &
            $1/6$      &
            $0$ 
                  \\
            $u$      &
            R& 
            $\bf 3$      &
            $\bf 1$      &
            $2/3$      &
            $0$ 
                  \\  
            $d$      &
            R&
            $\bf 3$      &
            $\bf 1$      &
            $-1/3$      &
            $0$ 
                  \\
            $H$      &
            -- & 
            $\bf 1$      &
            $\bf 2$      &
            $1/2$      &
            $0$ \\
            \hline 
            $\psi$      &
            L &
            $\bf 3$      &
            $\bf 1$      &
            $-1/3$      &
            $k$ 
            \\
            $\psi^c$      &
            R & 
            $\bf 3$      &
            $\bf 1$      &
            $-1/3$      &
            $k$ 
            \\
            $\Sigma_a$      &
            -- & 
            $\bf 1$      &
            $\bf 1$      &
            $0$      &
            $k$ 
                  \\
            $S$      &
            --& 
            $\bf 1$      &
            $\bf 1$      &
            $0$      &
            $0$ 
                  \\
        \bhline{1 pt}
        \end{tabular}
        \caption{The matter contents and their gauge charges in a Nelson-Barr type (spontaneous $CP$ violating) model.} 
        \label{tab:matter contents}
    \end{table}

\begin{figure}[t]
    \centering
    \includegraphics[width=0.8\linewidth]{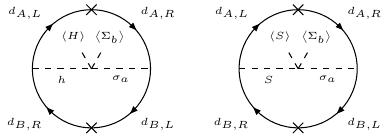}
    \caption{
        The two-loop diagrams that contribute to the radiative QCD theta parameter in the Nelson-Barr model. The external gluon fields are omitted. The cross ($\times$) represents the chirality flip. Subscripts $A$ and $B$ are for mass eigenstates of fermions.
        }
    \label{fig:NB_bubbles}
\end{figure}

It is pointed out by Ref.~\cite{Dine:2015jga} that the four-point scalar interactions of $H$ and $\Sigma_a$ lead to ${\rm arg\,det}({\cal M}_d)\ne 0$ at one-loop level and the QCD $\theta$ parameter is radiatively generated.\footnote{
When the four-point scalar interactions are absent, the QCD $\theta$ parameter is still generated at three-loop level. It is evaluated using the dimensional analysis in Refs.~\cite{Nelson:1984hg, Valenti:2021rdu}.}
It corresponds to the two-loop correction to the QCD  $\theta$ parameter, shown in the left diagram of Fig.~\ref{fig:NB_bubbles}.

 Furthermore, we introduce a real scalar field $S$ whose real vacuum expectation value leads to the singlet vector-like quark mass, $m$, as $m=f \langle S\rangle $ with a Yukawa coupling $- {\cal L} = f S \bar{\psi} \psi^c + {\rm h.c.}$. 
From a viewpoint of model-building,  however, the origin of $m$ should be related to the vacuum expectation values of $\Sigma_a$ since the observed CKM $CP$ phase of $O(1)$ is realized only when $m\sim |\xi_i|=|g^{ai}\langle \Sigma_a\rangle|$.
It is found that the four-point scalar coupling of $S$ and $\Sigma_a$ also generates the QCD  $\theta$ parameter at two-loop level, denoted as a right diagram of Fig.~\ref{fig:NB_bubbles}.
Then, we consider the following four-point scalar interactions,
\begin{eqnarray}
  V(\Sigma_a, H)&=&\gamma_{ab}\Sigma_a^\star \Sigma_b |H|^2+\frac12 \tilde{\gamma}_{ab}\Sigma_a^\star \Sigma_b S^2\cdots\,.    
\end{eqnarray}
We evaluate the QCD $\theta$ parameter at two-loop level assuming $\gamma_{ab}$ and $\tilde{\gamma}_{ab}$ are non-vanishing.
  
The down-type quark mass matrix is diagonalized as 
\begin{eqnarray}
{\cal M}_d&=&
O_d\bar{{\cal M}}_d U^\dagger_d\,,
\end{eqnarray}
where $\bar{{\cal M}}_d$ is the diagonal mass eigenvalue matrix
$\bar{{\cal M}}_d={\rm diag}(\bar{m}_{d_A})$ ($A=1,\cdots,4)$.  The inverse of ${\cal M}_d$ is given by the unitary matrices as 
\begin{eqnarray}
{\cal M}^{-1}_d&=&
U_d\bar{{\cal M}}^{-1}_d O_d^\dagger\nonumber\\
&=&\left(
\begin{array}{cc}
y_d^{-1} \langle H\rangle^{-1}&0\\
-\frac{1}{m}\xi^\dagger y_d^{-1} \langle H\rangle^{-1}&\frac{1}{m}
\end{array}
\right)\,.
\label{inveseofm}
\end{eqnarray}
We assume that $\langle H \rangle\ll m$, $|\xi_k|$. In this case, the unitary matrices $U_d$ and $O_d$ are given as\footnote{
This diagonalization comes from a mathematical identity \cite{Valenti:2021rdu}, 
\begin{eqnarray}
{\cal M}_d\tilde{U}_d&=&
\left(\begin{array}{cc}
    Y_d\langle {H}\rangle& Y\langle {H}\rangle\\
    0& M
  \end{array}
\right)\,.
\nonumber
\end{eqnarray}
}
\begin{eqnarray}
U_d=\tilde{U}_d\left(\begin{array}{cc}u_d&0\\0&1\end{array}\right)\,,&&
O_d=\tilde{O}_d\left(\begin{array}{cc}o_d&0\\0&1\end{array}\right)\,.
\label{udod}
\end{eqnarray}
Here,
\begin{eqnarray}
  \tilde{U}_d=
  \left(\begin{array}{cc}
    1-\frac{\xi\xi^\dagger}{|\xi|^2}(1-\frac{m}{M})&\frac{\xi}M\\
    -\frac{\xi^\dagger}{M}&\frac{m}{M}
  \end{array}
  \right)\,,&&
 \tilde{O}_d=
  \left(
  \begin{array}{cc}
    1&\frac{Y\langle H\rangle}{M}\\
    -\frac{Y^\dagger\langle H\rangle}{M}&1
  \end{array}
  \right)\,,
\end{eqnarray}
where $M^2=m^2+\xi^\dagger \xi$ and $Y^i=y_d^{ij}\frac{\xi_j}{M}=Y_d^{ij}\frac{\xi_j}{m}$.
The mass eigenvalues of the ${\cal M}_d$ are given by $\bar{m}_{d_4}= M$ and eigenvalues 
of $Y_d^{ij}\langle H\rangle$ for three SM down quarks,
\begin{eqnarray}
  Y_d^{ij}&=&y_d^{ik}\left[1-\frac{\xi\xi^\dagger}{|\xi|^2}\left(1-\frac{m}{M}\right)\right]_{kj}.\,
\end{eqnarray}
Two $(3\times 3)$ unitary matrices in Eq.~\eqref{udod},   $u_d$ and $o_d$, are for diagonalization of $Y_d$.

Now we evaluate the corrections to the QCD $\theta$ parameter at two-loop level, as shown in Fig.~\ref{fig:NB_bubbles}. They come from mixings between $H$ and $\Sigma_a$ and also between $S$ and $\Sigma_a$. Assuming $\gamma_{ab}$ and $\tilde{\gamma}_{ab}$ much less than one, we keep only the leading terms in the perturbation. In the case of diagrams in Fig.~\ref{fig:NB_bubbles}, they give the corrections to the QCD $\theta$ parameter as 
\begin{eqnarray}
\Delta\theta_\Sigma&=&
\frac{1}{8\pi^2}
y_d^{ij}g^{ak}\gamma_{ab} 
{\rm Im} \left[
[U_d]_{kA}\bar{m}_{d_A}
[O^\dagger_d]_{Ai}
[U_d]_{jB}\bar{m}_{d_B}
[O^\dagger_d]_{B4}
\right]
\nonumber\\
&&\times 
\frac{\langle H\rangle \langle \Sigma_b\rangle}
{m_h^2-m_{\Sigma_a}^2}
\left[
\left(I_{(2;2)}+\bar{I}_{(1;3)}+\bar{I}_{(3;1)}\right)(\bar{m}_{d_A}^2;\bar{m}_{d_B}^2;m_h^2)
-(m_h^2\rightarrow m_{\Sigma_a}^2)
\right]\,,\\
\Delta\theta_S&=&
\frac{1}{8\pi^2}
f g^{ak}\tilde{\gamma}_{ab} 
{\rm Im} \left[
[U_d]_{kA}\bar{m}_{d_A}
[O^\dagger_d]_{A4}
[U_d]_{4B}\bar{m}_{d_B}
[O^\dagger_d]_{B4}
\right]
\nonumber\\
&&\times 
\frac{\langle S\rangle \langle \Sigma_b\rangle}
{m_S^2-m_{\Sigma_a}^2}
\left[
\left(I_{(2;2)}+\bar{I}_{(1;3)}+\bar{I}_{(3;1)}\right)(\bar{m}_{d_A}^2;\bar{m}_{d_B}^2;m_S^2)
-(m_S^2\rightarrow m_{\Sigma_a}^2)
\right]\,,
\end{eqnarray}
where $m_h$, $m_S$, and $m_{\Sigma_a}$ are masses of 
$H$, $S$, and $\Sigma_a$, respectively. These results still include
fake IR bad behavior due to light quarks such as $1/\bar{m}_{d_k}$. However, using $\bar{I}_{(3;1)}(x_1,x_2,x_3) \simeq -1/(2 x_1)F_0(0,x_2,x_3)$ for $x_1\ll x_3$ and 
inverse matrix of ${\cal M}_d$ in Eq.~(\ref{inveseofm}), the IR behavior can be removed.  Then, when $\langle H \rangle\ll m$, $|\xi_k|$, the above results are reduced as 
\begin{eqnarray}
\Delta\theta_\Sigma&=&
-\frac{1}{16\pi^2}
\gamma_{ab} g^{ak}g^{ck}{\rm Im} [ \langle \Sigma_b\rangle
\langle \Sigma_c\rangle^\star]
\nonumber\\
&&\times 
\frac{1}
{m_h^2-m_{\Sigma_a}^2}
\left[
F_0(0;M^2;m_h^2)-(m_h^2\rightarrow m_{\Sigma_a}^2)
\right]\,,\\
\Delta\theta_S&=&
\frac{1}{8\pi^2}
\tilde{\gamma}_{ab} 
 g^{ak}g^{ck}{\rm Im} [ \langle \Sigma_b\rangle
\langle \Sigma_c\rangle^\star]\textbf{}
\nonumber\\
&&\times 
\frac{Mf \langle S\rangle}
{m_S^2-m_{\Sigma_a}^2}
\left[
\left(I_{(2;2)}+2 \bar{I}_{(3;1)}\right)(M^2;M^2;m_S^2)
-(m_S^2\rightarrow m_{\Sigma_a}^2)
\right]\,.
\end{eqnarray}
Since the $CP$ symmetry is spontaneously broken, the corrections to the QCD $\theta$ parameter are UV finite. It is found that $\Delta \theta_{\Sigma}$ comes mainly from the correction to light quark masses since it is given by the function $F_0$. On the other hand, $\Delta \theta_{S}$ is generated by the vector-like quark loop.  When $M$ is comparable to $m_S$ and/or $m_{\Sigma_a}$, the correction is not dominated by the correction to the vector-like quark mass, and the full evaluation of the two-loop diagram is required to evaluate them.

\section{Conclusion and Discussion} 
\label{sec:conclusion}

The QCD $\theta$ parameter is generated radiatively in spontaneous $CP$ or $P$ symmetry-breaking models, which solve the strong $CP$ problem. 
We scrutinized the QCD $\theta$ parameter at the two-loop level analysis. 
In the simplified models with $CP$-violating Yukawa interactions, we observed that the two-loop calculation of the radiative QCD $\theta$ parameter using the Fock-Schwinger gauge method is consistent with the effective field theory approach at the low-energy scale. 
Furthermore, we clarified the application scope of the Fujikawa method. 
When there is a scale hierarchy in the particle masses in $CP$-violating sector, the Fujikawa method is sufficient for evaluating the QCD $\theta$ parameter.
On the other hand, in the case of a small hierarchy, the Fujikawa method does not cover full contribution.
The Nelson-Barr model is an example that the Fujikawa method cannot evaluate the radiatively generated QCD $\theta$ parameter correctly. 
If the vector-like quark and additional scalars have comparable masses, the Fock-Schwinger gauge method should be used to evaluate the radiative QCD $\theta$ parameter.

It is an important subject to evaluate the three-loop contributions to the QCD $\theta$ parameter in some well-motivated models for the strong $CP$ problem, such as the left-right models, and to compare the predictions with the experimental bounds on the QCD $\theta$ parameter. In the evaluation, the effective theory approach would be useful if the CP-violating interaction can be integrated out.

%=======================================================
%        ACKNOWLEDGEMENTS
%=======================================================
\acknowledgments

This work is supported by the JSPS Grant-in-Aid for Scientific Research Grant No.\,20H01895 (J.H.) and No.\,21K03572 (J.H.).
The work of J.H.\ is also supported by 
World Premier International Research Center Initiative (WPI Initiative), MEXT, Japan.
This work is also supported by 
JSPS Core-to-Core Program Grant No.\,JPJSCCA20200002. 
This work was financially supported by JST SPRING, Grant Number JPMJSP2125. The author (N.O.) would like to take this opportunity to thank the ``Interdisciplinary Frontier Next-Generation Researcher Program of the Tokai Higher Education and Research System.''

\appendix

%=======================================================
%=======================================================

\section{Loop functions}
\label{app:loop}
    We define the loop functions in this section. The two-loop functions used in this paper are given as
    \begin{align}
      &I_{(n_1,\cdots;m_1,\cdots)}(x_1,\cdots;x_2,\cdots;x_3) \nonumber \\
      &\quad =(16\pi^2\mu^{2\epsilon})^2\int \frac{d^dp}{(2\pi)^{d}}\frac{d^dq}{(2\pi)^{d}}
       \frac{1}{[p^2-x_1]^{n_1}\cdots[q^2-x_2]^{m_1}\cdots[(p+q)^2-x_3]}\,,
       \label{eq:Idef}
    \end{align}  
    where the dimensional regularization is used on $d=4-2\epsilon$ dimension and $\mu$ is the renormalization scale.
    The functions can also be derived by derivative or finite difference of $I(x_1;x_2;x_3)$ ($\equiv I_{(1;1)}(x_1;x_2;x_3)$)
    as
    \begin{eqnarray}
      I_{(n;m)}(x_1;x_2;x_3)&=&\frac{1}{(n-1)!(m-1)!}
      \frac{d^{n-1}}{dx_1^{n-1}}\frac{d^{m-1}}{dx_2^{m-1}}
      I(x_1;x_2;x_3)\,,\\
      I_{(1,1;1)}(x_1,x_1^\prime;x_2;x_3)&=&\frac{1}{x_1-x_1^\prime}\left[
      I(x_1;x_2;x_3)-I(x_1^\prime;x_2;x_3)\right]\,,
    \end{eqnarray}  
    where
    \begin{eqnarray}
      I(x_1;x_2;x_3)&\equiv &I_{(1;1)}(x_1;x_2;x_3)\nonumber\\
      &=&(16\pi^2\mu^{2\epsilon})^2\int \frac{d^dp d^dq}{(2\pi)^{2d}}
        \frac{1}{[p^2-x_1][q^2-x_2][(p+q)^2-x_3]}\,.
    \end{eqnarray}
    The explicit form of
    $I(x_1;x_2;x_3)$ is given as
    \begin{eqnarray}
      I(x_1;x_2;x_3)&=&
    \bar{I}_\epsilon(x_1;x_2;x_3)+\bar{I}(x_1;x_2;x_3)\,,
    \end{eqnarray}
    with the UV divergent part
    \begin{align}
    \bar{I}_\epsilon(x_1;x_2;x_3)=-\sum_{i=1,2,3}x_i\left[
      \frac{1}{2\epsilon^2}-\frac{1}{\epsilon}\left(\log\frac{x_i}{Q^2}-\frac32\right) + \left(1+\frac{\pi^2}{12}-\log\frac{x_i}{Q^2}+\frac12 \log^2\frac{x_i}{Q^2}\right)\right]\,,
    \end{align}
    while the finite part 
     \begin{align}
     & \bar{I}(x_1;x_2;x_3)
      =
      -\frac12\biggl[(-x_1+x_2+x_3)\log\frac{x_2}{Q^2}\log\frac{x_3}{Q^2}
    +(x_1-x_2+x_3)\log\frac{x_1}{Q^2}\log\frac{x_3}{Q^2}
    \nonumber\\
    &\quad 
    +(x_1+x_2-x_3)\log\frac{x_1}{Q^2}\log\frac{x_2}{Q^2}
    -4\left(x_1\log\frac{x_1}{Q^2}
    +x_2\log\frac{x_2}{Q^2}
    +x_3\log\frac{x_3}{Q^2}\right)
    \nonumber\\
    &\quad 
    +5(x_1+x_2+x_3)+\xi(x_1,x_2,x_3)\biggr]\,,
   \\
   & \xi(x_1,x_2,x_3)=
    R\left[2 \log\left(\frac{x_3+x_1-x_2-R}{2 x_3}\right)\log\left(\frac{x_3-x_1+x_2-R}{2 x_3}\right)
    -\log\frac{x_1}{x_3}\log\frac{x_2}{x_3}
    \right.\nonumber\\
    &
   \quad  \left.
    -2{\rm Li}_2\left(\frac{x_3+x_1-x_2-R}{2 x_3}\right)
    -2{\rm Li}_2\left(\frac{x_3-x_1+x_2-R}{2 x_3}\right)
    +\frac{\pi^2}{3}
    \right]\,,
    \end{align}
    where $R=\sqrt{x_1^2+x_2^2+x_3^2-2 x_1 x_2-2x_2 x_3-2 x_3 x_1}$ and
    $Q^2\equiv4\pi \mu^2 {\rm e}^{-\gamma_E}$ \cite{Ford:1992pn,Espinosa:2000df,Martin:2001vx}. 
    Note that although the last terms of $\bar{I}_\epsilon$ are UV finite, 
    they do not affect any physical quantity \cite{Martin:2018emo}, whose terms are suppressed by $\mathcal{O}(\epsilon)$ at one-loop level, while they are uplifted to  $\mathcal{O}(\epsilon^0)$ at two-loop level.

The radiative correction to the QCD $\theta$ parameter at two-loop level is proportional to
${I}_{(2;2)}(x_1;x_2;x_3)+\bar{I}_{(3;1)}(x_1; x_2; x_3)+\bar{I}_{(3;1)}(x_2;x_1;x_3)
$
($\bar{I}_{(3;1)}$ is defined as Eq.~(\ref{eq:ibar31})). 
We provide it with some limits here. When $x_1\ll x_2,~x_3$, we have 
\beq
&  {I}_{(2;2)}(x_1;x_2;x_3)+\bar{I}_{(3;1)}(x_1;x_2;x_3)+\bar{I}_{(3;1)}(x_2;x_1;x_3)
\nonumber\\
 &\quad\quad =\frac{1}{2x_1}\left[1-\frac{1}{x_3-x_2}\left(x_3\log\frac{x_3}{\mu^2}-x_2\log\frac{x_2}{{\mu^2}}\right)\right]\nonumber\\
&\quad\qquad +\frac{1}{2x_2}\left[1-\frac{1}{x_3-x_2}\left(x_3\log\frac{x_3}{\mu^2}-x_2\log\frac{x_2}{\mu^2}\right)\right]\nonumber\\
&\quad\qquad +\frac{3x_3^2-4x_3 x_2+x_2^2+2 x_3^2 \log\frac{x_2}{x_3}}{2(x_3-x_2)^3}\log\frac{x_1}{x_3}\nonumber\\
&\quad\qquad -\frac{({x_2^2- 3x_2 x_3 -2x_3^2} ) \log\frac{x_2}{x_3}
-4x_3^2~{\rm Li}_2\left(1-\frac{x_2}{x_3}\right)}{2(x_3-x_2)^3}\,.
\eeq
To obtain this analytic formula, we expand $R$ by $x_1$,
\beq
R \simeq \pm \left[(x_3-x_2) 
- \frac{x_2 + x_3 }{x_3 - x_2} x_1 
- \frac{2 x_2 x_3 }{(x_3 - x_2)^3} x_1^2 - \frac{
 2 x_2 x_3 (x_2 + x_3) }{(x_3 - x_2)^5}x_1^3\right]\,,
\eeq
where the overall $+$\,($-$) sign is for $x_3 > x_2$ ($x_2 > x_3$).

When $x_1,~x_2 \ll x_3$, we obtain
\beq
&  
{I}_{(2;2)}(x_1;x_2;x_3)+\bar{I}_{(3;1)}(x_1;x_2;x_3)
+\bar{I}_{(3;1)}(x_2;x_1;x_3)\nonumber\\
 &\quad\quad =
 \frac{1}{2x_1}\left(1-\log\frac{x_3}{\mu^2}{ + \frac{x_2}{x_3}\log \frac{x_2}{x_3}}\right)
+\frac{1}{2x_2}\left(1-\log\frac{x_3}{\mu^2} { + \frac{x_1}{x_3}\log \frac{x_1}{x_3}} \right)
\nonumber\\
&\quad\qquad 
+\frac{2\pi^2+9 (\log\frac{x_1}{x_3}+\log\frac{x_2}{x_3})
  +6\log\frac{x_1}{x_3}\log\frac{x_2}{x_3}}{6x_3}\,,
\eeq
and
\beq
&  
{I}_{(2;2)}(x_1;x_1;x_3)+2\bar{I}_{(3;1)}(x_1;x_1;x_3)\nonumber\\
 &\quad\quad =
 \frac{1}{x_1}\left(1-\log\frac{x_3}{\mu^2}\right)
+\frac{\pi^2+{12} \log\frac{x_1}{x_3}
  +3\log^2\frac{x_1}{x_3}}{3 x_3}\,.
\eeq

%=======================================================
%        REFERENCES
%=======================================================
\bibliographystyle{utphys28mod}
\bibliography{cedm_theta}

\end{document}